\documentclass[]{emulateapj}
\usepackage{natbib,graphicx,amsmath,amsthm,ulem,color}
\pdfoutput=1
\usepackage{CJK}

\def\pomega{\varpi}

\newcommand{\be}{\begin{equation}}
\newcommand{\ee}{\end{equation}}
\newcommand{\beqn}{\begin{eqnarray}}
\newcommand{\eeqn}{\end{eqnarray}}
\newcommand{\bi}{\begin{itemize}}
\newcommand{\ei}{\end{itemize}}

\def\refnew#1{(\ref{#1})}
\def\erg{\, \rm erg}

\def\s{\, \rm s}

\def\yrs{\, \rm yrs}

\def\pomega{\varpi}

\usepackage{ulem}

\newcommand{\w}{}
\newcommand{\wyq}{}

\shorttitle{Diffusive Tidal Evolution}
\shortauthors{Yanqin Wu}

\begin{document}


\title{Diffusive tidal evolution for migrating hot Jupiters}


\author{Yanqin Wu  (武延庆)}
\affil{Department of Astronomy and Astrophysics, University of Toronto, Toronto, ON M5S 3H4, Canada}
\email{wu@astro.utoronto.ca}




\begin{abstract} 
  I consider a Jovian planet on a highly eccentric orbit around its
  host star, a situation produced by secular interactions with its
  planetary or stellar companions.  The tidal interactions at every
  periastron passage exchange energy between the orbit and the
  planet's degree-2 fundamental-mode. Starting from zero energy, the
  f-mode can diffusively grow to large amplitudes if its one-kick
  energy gain $ \geq 10^{-5}$ of the orbital energy.  This requires a
  pericentre distance of $\leq 4$ tidal radii (or $1.6$ Roche
  radii). If the f-mode has a non-negligible initial energy, diffusive
  evolution can occur at a lower threshold. The first effect can stall
  the secular migration as the f-mode can absorb orbital energy and
  decouple the planet from its secular perturbers, parking all
  migrating jupiters safely outside the zone of tidal disruption.  The
  second effect leads to rapid orbit circularization as it allows an
  excited f-mode to continuously absorb orbital energy as the orbit
  eccentricity decreases.  So without any explicit dissipation, other
  than the fact that the f-mode will damp nonlinearly when its
  amplitude reaches unity, the planet can be transported from a few AU
  to $\sim 0.2$ AU in $\sim 10^4$ yrs. Such a rapid circularization is
  equivalent to a dissipation factor $Q \sim 1$, and it explains the
  observed deficit of super-eccentric Jovian planets.  Lastly, the
  repeated f-mode breaking likely deposit energy and angular momentum
  in the outer envelope, and avoid thermally ablating the planet.
  Overall, this work boosts the case for forming hot Jupiters through
  high-eccentricity secular migration.
  \end{abstract}



\section{Introduction}

Hot Jupiters, the first-known population of extra-solar planets
\citep{MayorQueloz,Marcy}, orbit their stars at implausibly close
ranges, so close that they are not thought to have formed locally but have
been migrated inward, either by dynamical interactions with other
large bodies, or by gas in the protoplanetary disks
\citep{LinPapaloizou}. In the former scenario \citep[going by a number
of flavours, e.g., planet scattering, Kozai-Lidov migration, secular
chaos... ][]{Ford01,WuMurray, Nagasawa, WuLithwick}, angular momentum
exchanges between a Jovian planet, originally at a few AU, and its
neighbours (either stellar or planetary ones) gradually squeeze the
planet's orbit, causing it approach the star with an ever-decreasing
minimum distance.  Strong tides are raised on the planet whenever it
sweeps by its host star. It is hypothesized then that these tidal
sloshing can be dissipated into heat by friction inside the planet,
leading to orbital decay and circularization. A hot Jupiter is thus
born, as the direct result of tidal dissipation.

While neatly accounting for the presence of hot Jupiters and many of
their observed properties (e.g., their tight pile-up at a few
times the Roche radii, their lack of nearby-companions, the
metal-richness of their host stars...), theories of dynamical
migration all share three fatal weaknesses -- all related to the tidal
process. First, friction inside a gaseous planet like Jupiter has been
shown to be too weak, by orders of magnitude, to generate the required
dissipation \citep{GoldreichNicholson,Wu05b}.\footnote{ But see
\citet{OgilvieLin} for a success story in planets
  with large cores.}
Second, in numerical simulations, the angular momentum exchanges with
their secular perturbers oftentimes push these planets too close to
their stars, crucifying them in the process.
Hydrodynamics simulations found that if a planet comes inward of $2.7
r_t$, where the tidal radius $r_t = (M_*/M_p)^{1/3} R_p$, it will be
tidally disrupted within a handful of orbits \citep{Guillochon}. {\wyq
  Numerically, one finds that} up to $90\%$ of migrating hot Jupiters
can be pushed inward of this distance and go to waste
\citep{Petrovich,Munoz,Hamers17}.  As a result, the proposed
mechanisms fail to account for the observed frequency of hot
Jupiters. Third, in order to circularize the orbits, the tides need to
deposit inside the planet an amount of energy that is comparable to or
larger than the planet's binding energy. There is no guarantee that
any planet can survive this, rather than be thermally ablated.

Reviving previous investigations by
\citet{Mardling,Kochanek,IvanovPapaloizou}, I now consider a tidal
process for high eccentricity orbits. This process has the potential
to resolve all three of the above weaknesses.

At every periastron passage, tidal stretching and compression excite
oscillations inside the planet. Orbital energy is converted into fluid
motion, or, the orbital degree of freedom and the internal degrees of
freedom are coupled. The most important internal mode {\wyq for this}
is an $\ell=2$ f-mode. If one ignores the feedback from the mode to
the orbit, the orbit remains strictly periodic and
the internal mode is only excited to a finite (and typically small)
amplitude, much like that of a harmonic oscillator driven under a
periodic, non-resonant force
\citep{PT,LeeOstriker,Lai97,Smeyers91}. There is no long-term benefit
to this interaction. However, when the closest approach is small
enough, it is no longer valid to ignore the feedback. The oscillations
can acquire a sufficient amount of energy to alter the orbital period
significantly. \citet{Mardling,Kochanek} are the first to use
numerical simulations to show that the mode energy can now undergo
random-walk.
And \citet{IvanovPapaloizou} followed up by illuminating the
underlying physics. This goes as follows. At every passage, the f-mode
receives a kick from the tidal potential. The magnitude of this kick
can be considered roughly constant, as long as the peri-centre
distance is kept constant. Its phase, however, depends on the phase of
the f-mode pulsation at periastron. This in turn depends on the length
of an orbit, which is perturbed by the tidal energy exchange.  When
this phase is sufficiently random between kicks, the mode can be
launched into a random walk with its energy growing roughly linearly
in time.

In this work, I extend the result of \citet{IvanovPapaloizou} by
obtaining the quantitative criterion for diffusive tidal
evolution. This is performed for the case when the f-mode has zero
initial energy \citep{Mardling, IvanovPapaloizou,Dong}, and for the
case when the f-mode has a finite initial energy (\S
\ref{sec:exchange}). I give simple explanations for these criteria (\S
\ref{sec:conditions}) and consider the impacts of these physics on the
migration of hot Jupiters (\S \ref{sec:secular}).

\section{Energy exchange between Orbit and Mode}
\label{sec:exchange}

To consider the coupled evolution of the orbit and
the modes, I use the equations of motion first derived by
\citet{Lai97}, following that of \citet{PT}. An alternative
prescription, based on the variational principle, is derived by \citet{Gingold}.
I consider exclusively tides raised on the planet (mass $M_p$, radius
$R_p$) by the star (point mass $M_*$), ignoring effects of mode
dissipation (justified later).

I consider a core-less model for Jupiter, $R_p = 1.1 R_J$, $M_p = 1
M_J$. The slight size inflation mimics a young Jupiter on its cooling
contraction ($t \sim 1$ Gyrs).
Of most relevance is the period of the $\ell=2$ f-mode. Let us scale
the results of \citet{Gudkova}, $P_0 = 8502 \s$ for
Jupiter,\footnote{A simpler calculation, using the Cowling
  approximation, would have produced a mode period that is $\sim 50\%$
  shorter. That is not accurate enough.}  by $(R_p/R_J)^{2.1}$
\citep{Burrows} to obtain $P_0 = 1.04\times 10^4 \s$.

I first focus on the one-kick energy, the amount of energy an f-mode
would acquire after one periastron passage, if it has zero initial
energy. This is demonstrated to affect the behaviour of the f-mode
when multiple passages are considered. Lastly, it is shown that the
initial energy of an f-mode also affects the dynamics.

\subsection{Equations of Motion}
\label{subsec:eos}


Let ${\bf D}$ be the vector from the centre of the planet to the star.
Written in spherical coordinates in the co-moving frame of the planet,
${\bf D} = [D(t), {\pi/2},\Phi(t)]$. The star (mass $M_*$) exerts a
tidal potential $U({\bf r},t)$ for fluid at position ${\bf r}$ inside
the planet
and excites motion. I decompose the excited motion, expressed in
displacement vector, as ${\bf \xi} = \sum_\alpha a_\alpha(t) {\bf
  \xi}_\alpha({\bf r})+c.c.$, where ${\bf \xi}_\alpha$ is the
eigenvector for eigenmode $\alpha$, $a_\alpha$ its complex amplitude,
and $\omega_\alpha$ its real frequency (dissipation ignored). {\w As
  the planet is axis-symmetric, the eigenfunctions can be decomposed in
  the azimuthal direction into periodic functions (i.e.,
  $\cos(m\phi)$). And in the following, we adopt the sign convention
  of $e^{i \sigma t + i m \phi}$, so a positive $m$ indicates a
  retrograde mode in the inertial frame,\footnote{The sense in the
    planet's rotating frame depends on the direction of the planet's
    spin.} while a negative value that of a prograde one.} Here $c.c.$
stands for complex conjugate.  The eigenfunction is normalized as
$\int d^3x \rho {\bf\xi}_\alpha \cdot {\bf \xi}^*_\alpha = M_p
R_p^2$. As a result, all perturbed quantities have natural dimensions
(e.g., $\xi$ has the dimension of length, and $a_\alpha$ is
dimensionless).  The amplitude of the normal mode is excited by
the tidal potential as
\begin{eqnarray}
{\ddot{a}}_{\alpha}& = & - \omega_\alpha^2 a_\alpha + \sum_\ell
{{G M_* W_{\ell m } Q_{\alpha\ell}}\over{M_p R_p^2 D^{\ell+1}}} e^{-i m \Phi}
\nonumber \\
& = & - \omega_\alpha^2 a_\alpha +  {{G M_*}\over {D^3}} \times 
\left({R_p}\over{D}\right)^{\ell-2}\, 
{ W_{\ell m } Q'_{n\ell}} e^{-i m \Phi}\, .
\label{eq:modeamplitude}
\end{eqnarray}
Here, I define a dimensionless form of
the tidal integral as
\begin{equation}
 Q'_{n\ell}  =  {{Q_{n\ell}}\over {M_p R_p^{\ell}}} =
 {1\over{M_p}} \, \int \rho r^2\, 
 \left({r\over{R_p}}\right)^\ell \left({{\delta \rho}\over{\rho}}\right)_{n\ell}
 \, dr \, .
\label{eq:defineQ'}
\end{equation}
where $\delta \rho$ is the Lagrangian density perturbation and is
related to the displacement ${\bf \xi}_\alpha$ by the equation of mass
conservation.  Definitions for the geometry factor $W_{\ell m}$ and
the tidal overlap integral $Q_{n\ell}$ are given in \citet{PT}. In our
case, the overlap is nonzero only between the $\ell$-term of the tidal
potential and a mode with degree $\ell$.
Moreover, the dimensionless factor $Q'_{n\ell}$ does not depend on
planet mass or radius, and is of order unity for f-modes of all
spherical degree $\ell$. Numerically, I find $Q'_{n\ell} \approx
0.5$ for f-modes.

In the mean time, the gravitational moment of the excited mode acts on
the orbit. Together with
the monopole potential ($ - GM_* M_p/D$), this moves the orbit as
\begin{eqnarray} {\ddot D} & = & D {\dot\Phi}^2 - {{G(M_*+M_p)}\over
    D^2} \times \nonumber \\
  & & \left[ 1 + \sum_\alpha {(\ell+1)} \,
    \left({{R_p}\over{D}}\right)^\ell\,
    {W_{\ell m} Q'_{n\ell}}(a_{\alpha}e^{i m \Phi}  + c.c.)\right] \nonumber \\
  {{d(D^2 {\dot \Phi})}\over{dt}} & = & {{G(M_*+M_p)}\over
    D} \times \nonumber \\
  & & \sum_\alpha i m \left({{R_p}\over{D}}\right)^\ell {W_{\ell m}
    Q'_{n\ell}} (a_{\alpha}e^{i m \Phi} + c.c.)\, .
\label{eq:orbitevolve}
\end{eqnarray}

Physically, eq. \refnew{eq:modeamplitude}-\refnew{eq:orbitevolve} can
be thought of as describing the interactions between two coupled
harmonic oscillators (the mode and the orbit). When the coupling
strength (tidal interaction) is weak, the oscillators exchange energy
periodically, with no long term effect; when it is strong, the
exchange is ergodic and drives the system toward energy equi-partition
between the two oscillators.

The total energy of the system should remain constant at all times, 
\begin{eqnarray}
E_{\rm tot} & = &  E_{\rm orb} + E_{\rm mode}  + V_{\rm tide}\nonumber \\
& = & - {{G 
    M_*M_p}\over{D}} + {1\over 2} \mu \left[{\dot D}^2 + (D {\dot
    \Phi})^2\right] \nonumber \\
& &  + \sum_\alpha \left( {\dot a_\alpha} {\dot a_\alpha}^* +
  \omega_\alpha^2 a_\alpha a_\alpha^* \right) M_p R_p^2\nonumber \\
& & - \sum_\alpha {{G M_*M_p}\over D}
\left({{R_p}\over{D}}\right)^\ell  {W_{\ell m}
    Q'_{n\ell}} \left[a_{\alpha} e^{im\Phi} + c.c.\right]\, .
\label{eq:Etot}
\end{eqnarray}
Here, $\mu = M_* M_p/(M_*+M_p)$ is the reduced mass, and the last
term represents the interaction energy between oscillations and
stellar gravity.
In practice, the conservation of total energy is used to ascertain the
accuracy of our numerical procedure.

Numerical integration of this system requires special attention.  With
the usual Runge-Kutta technique, the energy error grows rapidly and
becomes intolerable after just a few orbits. This is caused both by
the rapid mode oscillation and by the extremely small time-step
required for a highly eccentric orbit. I construct a special
integrator that is analogous to the drift-kick-drift symplectic orbit
integrator \citep{WisdomHolman} for planetary dynamics. In the drift
phase, the eccentric orbit and the oscillation mode are each advanced
forward in time analytically, assuming no interaction; and in the kick
phase, they are advanced by the amount of mutual interaction
integrated over the entire time-step. This strategy avoids some
numerical instabilities, but it still requires a very small time-step
($dt \sim 0.001 P$) to ensure satisfactory energy conservation.

\subsection{One-Kick Energy}
\label{subsec:onepass}

An important quantity in this problem is the amount of energy imparted
to a mode after one periastron passage, assuming initially zero
amplitude.  This quantity reflects the strength of tidal interaction
and is later used to separate the long-term evolution into two
regimes. Here, I compare the analytical expression for this quantity
\citep[first derived by][]{PT} against results of numerical integration.

The energy gain for mode $\alpha$ is \citep[][confirmed for our
normalization and complex notation]{PT}
\begin{eqnarray}
  \Delta E_{\rm kick} & = &  \int dt \int d^3 x \rho {{\partial {\bf \xi_\alpha}}\over{\partial
      t}} \cdot \nabla U \nonumber  \\
  & = & 4 \pi^2 M_p R_p^2 \left({{GM_*}\over{D_p^3}}\right)^2
  \left({{R_p}\over{D_p}}\right)^{2\ell - 4} |Q_{n\ell}'|^2 \, 
  |K_{\ell m} (\omega_\alpha)|^2 \nonumber \\
  & = & {\w 4}
\pi^2 M_p R_p^2 
  \left({{R_p}\over{D_p}}\right)^{2\ell - 4} |Q_{n\ell}'|^2 \, 
{{  |K_{\ell m} (\omega_\alpha)|^2}\over{T_{\rm peri}^4}} \, ,
\label{eq:Ekick}
\end{eqnarray}
where the periastron distance $D_p = a(1-e)$, and $T_{\rm peri}$ is
the timescale of periastron passage,
\begin{equation}
T_{\rm  peri} \equiv \sqrt{{{D_p^3}\over {GM_*}}}\approx 
1.41\times 10^4 {\rm s}\, 
\left({{D_p}\over{0.02 {\rm AU}}}\right)^{3/2} 
\left({{M_*}\over{M_\odot}}\right)^{-1/2}\, .
\label{eq:omegaperi}
\end{equation}
The orbit integral $K_{\ell m}$, 
\begin{equation}
K_{\ell m}  (\omega)\equiv {{W_{\ell m}}\over{2 \pi}} \int
dt \left({{D_p}\over{D(t)}}\right)^{\ell+1} \exp{i \left[\omega t + m
  \Phi(t)\right]}\, ,
\label{eq:defineK}
\end{equation}
quantifies how well the time-varying tidal potential is interacting
with the time-varying oscillation {\w over one orbit} and has the dimension of time.
Contribution to this integral arises mostly when $D(t) \sim D_p$, over
a duration $T_{\rm peri}$.
So if we write $K_{\ell m} = f T_{\rm peri}$, using the dimensionless
factor $f$ to account for both the geometry, and more importantly, the
cancellation in the integrated tidal forcing arising from the fact
that the mode may oscillate multiple cycles during a single periastron
passage, we find $f = 0.006$ for our $\ell=2$ {\w , $m=-2$ (prograde)}
f-mode at $D_p = 0.02$ AU, {\w and some $3000$ times smaller
  for the retrograde mode, assuming zero spin.}  So from now on, I
will only focus on the $\ell=2, m=-2$ {\w prograde} mode and drop the
mode subscript $\alpha$ accordingly.

The $f$-factor drops off exponentially for modes with shorter periods.
This excludes all but the longest period {\w (i.e., lowest degree)}
f-mode as being relevant for tidal interaction. It also suggests that,
if the {\w $m\neq 0$} f-mode is shifted to a longer period by planet
spin, the strength of tidal interaction increases.

\begin{figure} 
\includegraphics[width=0.49\textwidth,trim=0 420 30 80,clip=]{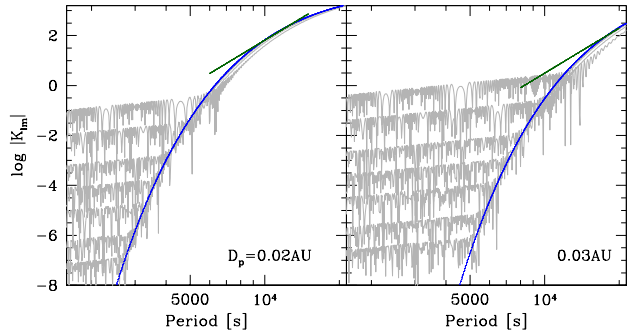} 
\caption{The logarithm of the orbital integral ($K_{\ell m}$) for the
  $\ell=2, m=-2$ mode, when the mode period takes on a range of values
  (horizontal axis, in second). The elliptical orbits take on $D_p =
  0.02$ AU (left panel) and $0.03$ AU (right panel). 
  The seven grey curves in each panel correspond to orbits with a
  range of semi-major axis, $a = 10, 5, 2, 1, 0.5, 0.2, 0.1$ AU (from
  bottom to top).  The numerical results agree with the analytical
  expression for parabolic orbits (blue lines, eq. \ref{eq:dong}) at
  large $a$ (highly eccentric), but deviate from it when $a$ is
  lower. This is more severe at the right hand panel.  The short green
  lines represent our simple power-law fit (eq. \ref{eq:Klmwu}). }
\label{fig:Klmfits-new}
\end{figure}

For a parabolic orbit, \citet{Lai97} provided an analytical expression
for $K_{2,-2}$ as 
\be K_{2,-2}
(\omega) = {{2 z^{3/2} e^{-{2\over3}
      z}}\over{\sqrt{15}}} \left(1 - {{\sqrt{\pi}}\over{4
      \sqrt{z}}}\right)\, T_{\rm peri} \, ,
\label{eq:dong}
\ee where $z=\sqrt{2} \omega T_{\rm peri}$.  I compare this
expression with integration results using elliptical orbits
(Fig. \ref{fig:Klmfits-new}). For orbits with large semi-major axis
where the parabolic limit is more appropriate, this expression is
reproduced. Importantly, the forcing strength depends only on the
value of $D_p$, not on the actual shape of the orbit ($a$, $e$). This
arises because highly eccentric orbits with the same periastron differ
little in geometry from the parabolic trajectory. This independence
allows the f-mode to continuously absorb energy as the orbit is being
circularized ($D_p$ remains roughly constant).  In the range of $z$
that is of interest to us, one can further simplify the above
expression into a power-law,
\begin{equation}
K_{2,-2} \approx 1.79\times 10^4 z^{-6}\, T_{\rm peri}\,  .
\label{eq:Klmwu}
\end{equation}

In contrast, for orbits that are more circular (smaller $a$),
numerical results show that $K_{\ell m}$ generally lies above
eq. \refnew{eq:dong} and exhibits many resonance features. This is
because contribution to the orbit integral is no longer strictly
limited to from near the periastron.  In the limit that the orbit is
circular, the entire orbit contributes and $K_{\ell m}$ is dominated
by resonances for which the mode period is an integer fraction of the
orbital period.  In this work, I will focus on the regime where the
parabolic expression is valid. There may be interesting dynamics in
the resonant regime.

We are interested in the fractional energy absorption, one that is {\w
  scaled by} the orbital energy. Defining $E_0 = - E_{\rm tot} - G M_*
M_p/2a_0$, we write
\begin{eqnarray}
{{\Delta E_{\rm kick}}\over{E_0}} 
& \approx & 1.3 \times 10^{-5} \, \left({a_0\over{1 {\rm AU}}}\right) 
\left({{D_p}\over{0.02 {\rm AU}}}\right)^{-3} 
\left({{R_p}\over{1.1 R_J}}\right)^{2}\, \nonumber \\
& & 
\times \left[{{Q_{n\ell}'}\over{0.5}} \right]^2 
\left[{{K_{\ell m} (\omega)}\over{0.006 T_{\rm peri}}}\right]^2\, .
\label{eq:fracE}
\end{eqnarray}
This quantity drops with increasing $D_p$ very steeply -- adopting the
rough scaling for $K_{\ell m}$ as in eq. \refnew{eq:Klmwu}, $\Delta
E_{\rm kick}/E_0 \propto D_p^{-21}$. It also rises with mode period as
$P^6$.
I confirm the analytical expression for the one-kick energy using the
numerical integrator.

{\w A few words about the impact of planet spin.}  At slow rotation,
the Coriolis force perturbs the frequency of an $m\neq 0$ mode away
from that of the $m=0$ mode as $\omega_m = \omega_0 - m (1-C_{n\ell})
\Omega_s $, where $\Omega_s$ is the spin rate and the rotational
splitting integral $C_{n\ell}$ is
$C_{n\ell} = {1/{M_p R_p^2}} \int r^2 dr \rho (2 \xi_r \xi_h +
\xi_h^2)$. 
For our $\ell=2$ f-mode, $C_{02}= 0.48$, or $\omega_{-2} \approx
\omega_0 + \Omega_s$. {\wyq So, while in a non-rotating planet, the
  prograde mode couples to the tidal potential much more strongly,} in
  a planet with prograde (relative to the orbit) spin, the prograde
  mode is shifted to a higher frequency, leading to weaker tidal
  coupling; in the mean time, its retrograde counterpart now has a
  lower frequency and can couple more effectively to the tidal
  potential.  This leads to interesting interplay between the
  planetary spin and mode excitation, a dynamics discussed in detail
  in \citet{IvanovPapaloizou}.


\subsection{Diffusion I: $E_{\rm init} = 0$}
\label{subsec:examples}

\begin{figure} 
\includegraphics[width=0.45\textwidth,trim=0 180 20 110,clip=]{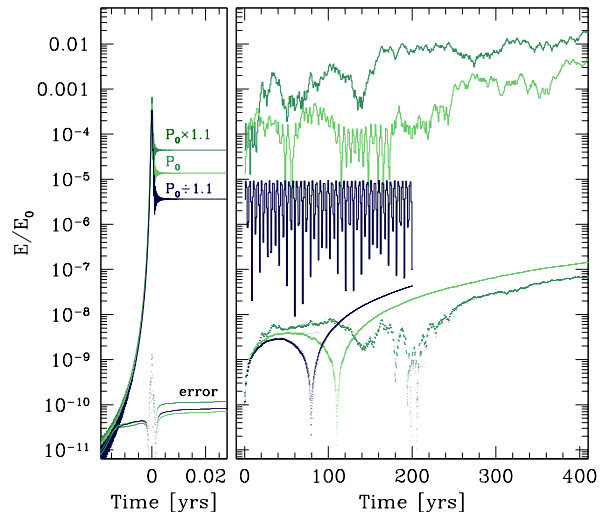} 
\caption{Tidal excitation over multiple passages for a Jovian planet
  initially on a $a=1$AU, $D_p=0.02$ AU orbit. The solid lines plot
  the fractional mode energy for the $\ell=2, m=-2$ f-mode when its
  period is varied from $10\%$ below to $10\%$ above its fiducial value
  ($P_0 = 1.04\times10^4 \s$), while the dots (lower group) show the
  fractional energy error in the numerical integrator.  The left panel
  focuses on the first passage (occurring at time zero), and the right
  panel that over multiple passages. Depending on the value of the
  one-kick energy, the long-term dynamics
  bifurcates into two behaviour:  lowly excited modes remain
  oscillating at a few times  their
  one-pass value; while highly excited modes can undergo random-walk in
  energy. The boundary between the two lies at $\Delta E_{\rm kick}/E_0 \sim 10^{-5}$.
}
\label{fig:example}
\end{figure}

Evolution after the first passage is studied numerically.  With the
special integrator, I am able to integrate the dynamics forward for
a satisfactory amount of time.  In Fig. \ref{fig:example}, I present
some results of such an integration.  The planet is on a highly
eccentric orbit of $a=1$AU, $e=0.98$ ($D_p = 0.02$AU). In this set-up,
since the initial orbital energy is of order the binding energy of the
planet, the f-mode reaches order unity amplitude (surface radial
displacement of order radius) when the mode energy reaches of order
$E_0$.

In each run, I consider a single $\ell=2, m=-2$ f-mode with a slightly
different period (ten percent around the fiducial period $P_0$) and
initially zero energy.  As is shown in Fig. \ref{fig:example}, after
the first passage, the mode acquires a different amount of energy that
rises with the f-mode period (eq. \ref{eq:fracE}).  And as one
continues to integrate the interactions, one sees that there is a
bifurcation in mode energy \citep{Mardling}: some exhibit
quasi-periodic oscillations in mode energies, with the maxima
comparable to or at most a few times larger than the one-kick value;
while mode energy in models with longer periods undergo random-walk
and rise to larger and larger values over time. As this occurs at the
price of the orbital energy, the orbit shrinks.  Meanwhile, on account
of the small moment of inertia of the planet, the f-modes do not
absorb a significant amount of the orbital angular momentum
\citep{IvanovPapaloizou}.  The latter is roughly conserved, with the
result that the pericentre distance remains largely constant during
the evolution. This in turns allows the f-modes to continue growing
unabatedly.
The bifurcation between the two behaviour appears to lie where the
fractional energy gain $\Delta E_{\rm kick}/|E_0| \sim 10^{-5}$.  This
will be explained in
 \S \ref{sec:conditions}.

Our numerical integrator guarantees energy conservation
(eq. \ref{eq:Etot}) to better than $10^{-10}$ over every single
passage, but as it is not symplectic, energy error does grow with
time. The fractional error reaches of order $10^{-7}$ after $\sim 400$
passages.  The integration is not be trusted when the energy error
becomes comparable to the mode energies, though for models that
undergo random-walk, this comes at a much later stage. For these
models, our results can be trusted to thousands of orbits and more.

\subsection{Diffusion II: $E_{\rm init} > 0$}
\label{subsec:example2}

The above bifurcation, for a f-mode with initially zero energy, has
been observed by \citet{Kochanek,Mardling,IvanovPapaloizou}. Here, I
report on a phenomenon that occurs for f-modes with some initial
energies, noted briefly previously by \citep{Mardling}.

\begin{figure}
\includegraphics[width=0.45\textwidth,trim=0 170 10 80,clip=]{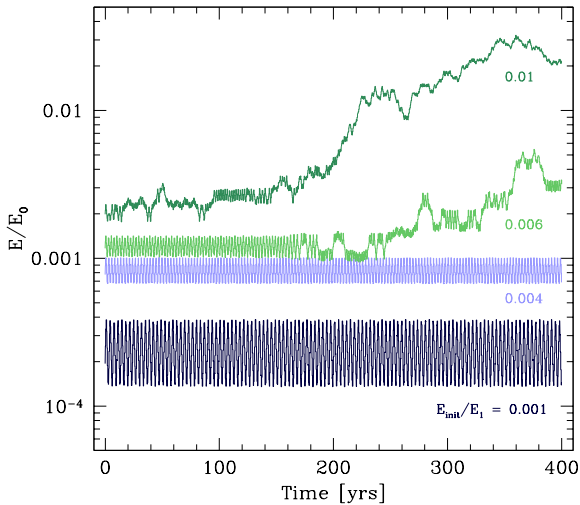}
\caption{The time evolution of mode energy (normalized by the orbital
  energy) for a planet at an orbit of $a=0.5$AU, $D_p = 0.025$
  AU.  The expected one-kick energy for this mode ($P= P_0  =
  1.04\times 10^4 \s$) is $\Delta E_{\rm kick}/E_0 \sim
  8\times 10^{-6}$. This is too weak to have allowed the mode to
  diffuse, if it starts with zero energy. However, as one endows
  the mode with an increasing amount of initial energy (values of
  $E_{\rm init}/E_1$ as marked), random-walk sets in again when 
$E_{\rm init}/E_1  \geq 0.005$, 
The rate of diffusion rises with the value of $E_{\rm init}$. }
\label{fig:example_nl}
\end{figure}

Fig. \ref{fig:example_nl} examines a mode that receives a weak
one-kick energy ($\Delta E_{\rm kick}/E_0 \sim 8\times 10^{-6}$). This
should not have undergone diffusion. However, I experiment with
endowing the mode with a varying amount of initial energy, quantified
by another energy unit, $E_1 \approx \sqrt{0.1 GM_p^2/R_p} = 3\times
10^{42} \erg$. This is the mode energy when its surface radial
displacement reaches unity. One finds that whenever $E_{\rm init}/E_1
\geq 0.005$, corresponding to a surface displacement of $\geq 7\%
R_p$, diffusion can set in again. This threshold corresponds to
$E_{\rm init}/E_0 \sim 8\times 10^{-4}$, or, the geometric mean of the
two energies of relevance, $(\Delta E_{\rm kick}/E_0)^{1/2} (E_{\rm
  init}/E_0)^{1/2} \sim 8\times 10^{-5}$.
This observation is explained below. 

\section{Conditions for F-mode Diffusion}
\label{sec:conditions}

Examples in \S \ref{subsec:examples}-\ref{subsec:example2} show that
there is a certain threshold of interaction for f-mode energy to
diffuse that depends on both the one-pass absorption, as well as the
initial energy in the f-mode. Here, I study the origin for these
thresholds using a mapping model that accurately describe the
physics. This approach was first invented by \citet{IvanovPapaloizou},
adopted in \citet{Dong}, and I develop it further here.

\subsection{Mapping and the Toy Model}

\begin{figure} 
\includegraphics[width=0.45\textwidth,trim=0 150 60 100,clip=]{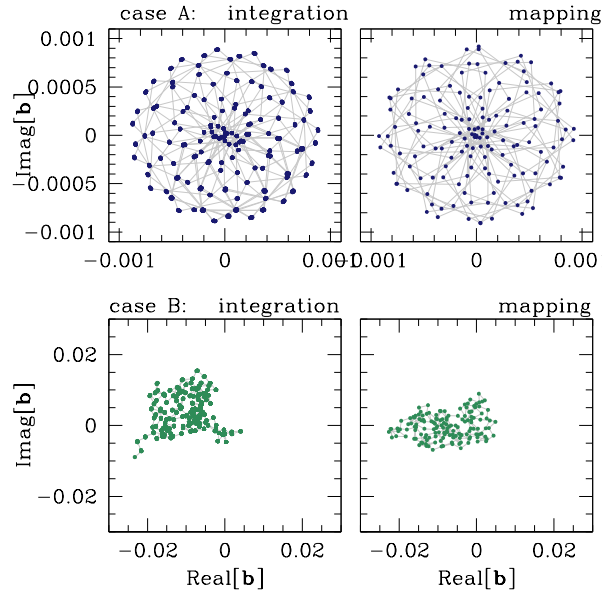}
\caption{Behaviour of the complex amplitudes (${\bf b}$, points),
  obtained using direct numerical integrations (left panels), and the
  simple toy-model (right panels), over $150$ passages.  The top case
  is for $a_0 = 1$AU, $D_p = 0.02$ AU and $P = P_0/1.1$, with $\Delta
  E_{\rm kick}/E_0 \sim 4 \times 10^{-6}$; while the bottom case is a
  mode with $P=1.1\times P_0$ that experiences a stronger kick
  ($\Delta E_{\rm kick}/E_0 \sim 5\times 10^{-5}$) and is launched
  into diffusion. The toy-model is physically accurate, because using
  the same parameters, it reproduces both behaviour correctly.}
\label{fig:randomwalk_dots}
\end{figure}

First, the tidal problem can be reduced to one of mapping.  The free
oscillation of the mode goes as $a(t) \propto \exp(i \omega t)$. So I
define a new complex mode amplitude to remove the rapid oscillation,
\begin{equation}
{b} = a (t) \exp(- i \omega t)\, .
\label{eq:balpha}
\end{equation}
This amplitude remains constant throughout most of the orbit when the
mode is freely oscillating, and is ``kicked'' by a discreet amount
when the planet passes through the periastron. Written in vector
form, 
\begin{equation}
{\bf b}_i = {\bf b}_{i-1} + \Delta {\bf b}_i\, ,
\label{eq:bj}
\end{equation}
where the complex increment from the $i$-th kick is $\Delta {\bf b}_i
= |\Delta \bf b_i| e^{i \Delta \theta_i }$.  The evolution is now
encapsulated in the vector addition of a discreet series of complex
amplitudes ${\bf b}_i$.  The left panels of
Fig. \ref{fig:randomwalk_dots} translate results of our numerical
simulations into such a mapping.

I now proceed to construct a simple toy-model that yields the same
mapping as the detailed numerics, one that is physically
accurate. First, let us consider the magnitude and phase of
individual kicks ($\Delta {\bf b}_i$).

The magnitude of individual kicks should be roughly constant in a
given system ($|\Delta {\bf b}_i| = |\Delta b|$). Observing
eq. \refnew{eq:modeamplitude}, one realizes that the kick only depends
on the orbital shape near periastron and the mode period. The latter
is roughly conserved during the evolution, as the mode-orbit
interactions do not absorb much of the orbital angular momentum,
thereby conserving $D_p$. In this study, I  take the mode period to be
constant.\footnote{This is only valid if one assumes that the f-mode,
  when it is dissipated, does not affect the planetary bulk structure
  and spin.}

The phase of the $i$-th kick, $\Delta \theta_i$, depends only on the
alignment between the pre-existing f-mode and the tidal potential at
the time of kicking, which in turn depends on the angle the mode has
rotated through in-between the kicks. Or, $\Delta \theta_i = {\rm
  Mod}(\omega \Delta T, 2\pi)$ with $\Delta T$ being the time between
the $i-1$-th and $i$-th passages.  

I now proceed to consider feedback onto the orbit. To produce a
simple toy-model, I set $\Delta T$ to be the instantaneous orbital
period. This then relates $\Delta T$ to the mode energy as,
\begin{equation}
\Delta T  =  {\w P_{\rm orb,0} \left({{E_{\rm orb}}\over{E_{\rm
        orb,0}}}\right)^{3/2}\, , }
\label{eq:toy}
\end{equation}
where {\w $E_{\rm orb} = E_0 + E$, with} $E_0 = GM_* M_p/2 a_0$, and
the mode energy $E= 2 \omega^2 |b|^2 M_p R_p^2$ (see
eq. \ref{eq:Etot}).  The mapping model is now complete.

In Fig. \ref{fig:randomwalk_dots}, it is shown that such a simple
model can accurately reproduce the outcomes of direct
integrations. One can now proceed to use this toy-model to efficiently
survey the parameter space, to determine how the diffusion threshold
depends on various parameters, and to explain its origin.


\subsection{Threshold I:  $E_{\rm init} = 0$}

\begin{figure} 
\includegraphics[width=0.48\textwidth,trim=0 0 10 0,clip=]{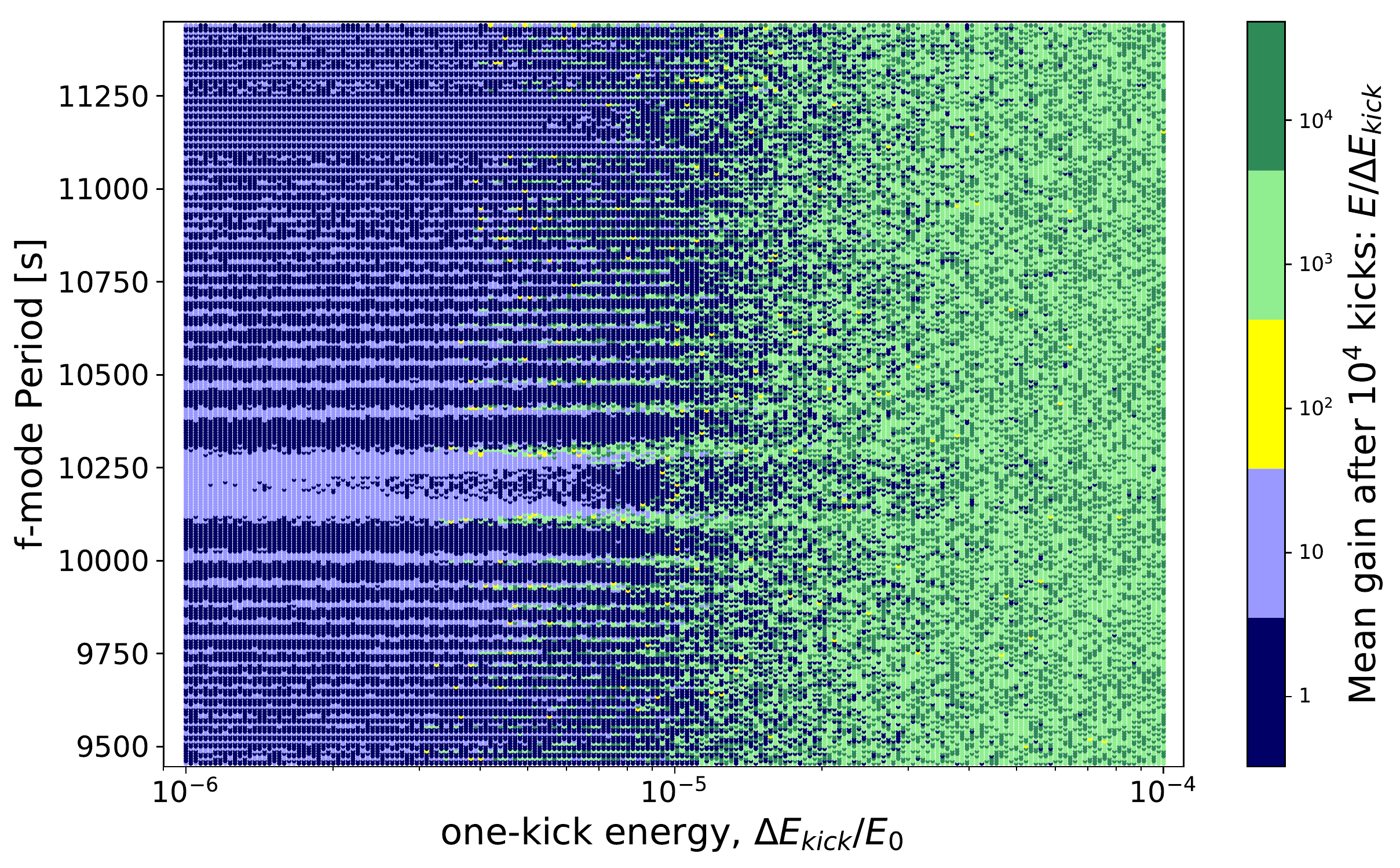} 
\caption{The toy-model results. Here, we scan through a range of mode
  periods (vertical axis) and kick magnitudes (fractional energy gain,
  horizontal axis), starting from zero amplitudes for the mode. The
  color at each point represents the mean magnification in mode energy
  (mean energy divided by the one-kick energy, color bar on the right)
  over $10^4$ passages.  Diffusive energy growth occurs when the kick
  magnitude is larger than about $10^{-5}$. Horizontal features
  correspond to modes with frequencies that are in resonance with the
  orbit. Here, the initial orbital period is 1 year (same as in
  Fig. \ref{fig:example}).}
\label{fig:randomkick-color}
\end{figure}

Starting from zero initial energy, the toy-model shows the same
bifurcation in mode growth as the direct integration.  Using the same
parameters as those in Fig. \ref{fig:example} ($a = 1$ AU, mode period
$P \sim P_0$), one finds the boundary to be also at $\Delta E_{\rm
  kick}/E_0 \sim 10^{-5}$ (Fig. \ref{fig:randomkick-color}).

What produces such a threshold? It turns out that even when the
fractional kick energy is a very small number, its effect on the kick
phase is not: it is amplified by the large number of oscillations in
an orbital period. Above the observed threshold, the variation in the
kick phase between successive kicks is, \be \delta (\Delta \theta_i)
\sim {3\over 2} \left(\omega P_{\rm orb,0} \right) {{\Delta E_{\rm
      kick}}\over{E_0}} \geq 0.3 {\rm radian}\, .
\label{eq:ddtheta}
\ee
This is now sufficiently large that the kicks can be considered to be
un-correlated in phase. As a result, 
\be
|{\bf b}_i|^2 = \sum_i |\Delta b|^2 + 2\sum_{i,j, i\neq j}
\Delta {\bf b}_i \cdot \Delta {\bf b}_j \approx \sum_i |\Delta b|^2\, .
\label{eq:randomb}
\ee Or, the mode energy grows at a roughly linear rate, \be {{dE}\over{dt}}
\approx {{\Delta E_{\rm kick}}\over{P_{\rm orb}}}\, .
\label{eq:diffusion}
\ee In contrast, weaker exchanges do not scramble the kick phases and
they are tightly correlated. Eq. \refnew{eq:toy} in this case acts as
a restoring potential that limits the mode energy to within a few
times the one-kick value.

The threshold energy depends on the mode period in a complicated
way. Orbital resonances may be partially responsible for this.  In the
following study, I adopt a threshold of
\begin{eqnarray}
{{\Delta E_{\rm kick}}\over{E_0}} & = & {1\over{2\omega
  P_{\rm orb}}} \nonumber \\
& = & 2\times 10^{-5}
  \left({{a_0}\over{1 {\rm AU}}}\right)^{-3/2}
  \left({{P}\over{1.04\times 10^4\s}}\right)\, ,
\label{eq:kickmin}
\end{eqnarray}
as a rough average. 



\subsection{Threshold II: $E_{\rm init} > 0$}
\label{subsec:findingII}

\begin{figure}
\includegraphics[width=0.36\textwidth,trim=0 170 100 100,clip=]{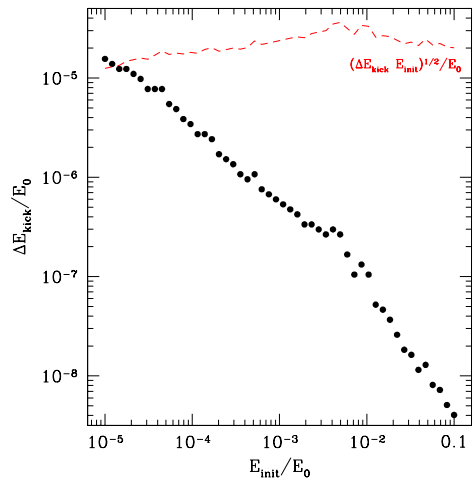}
\caption{The threshold one-kick energy for diffusion 
are plotted here in dots, as a function of the initial f-mode energy,
while the geometric
means of these two energies are shown as a red curve. 
 The threshold value is obtained using the toy mapping model and is 
defined as, above this one-kick energy, $50\%$
 of the system have diffused by more than 
$10^4 \Delta E_{\rm kick}$ from their initial energies, after $10^5$
passages.  Here, $P_{\rm orb,0} = 1$ yr, and the
mode period ranges from $P_0/1.1$ to $1.1\times P_0$. 
For comparison, the threshold kick at zero $E_{\rm init}$ lies at
$\Delta E_{\rm kick}/E_0 \sim 2\times 10^{-5}$. 
}
\label{fig:randomkick_finiteamp}
\end{figure}

Now consider the same problem but with an initial mode energy $E_{\rm
  init} \geq \Delta E_{\rm kick}$. Our toy-model shows that the
threshold kick is now much reduced and lies at (Fig. \ref{fig:example_nl}),
\begin{equation}
\sqrt{{{\Delta E_{\rm kick}}\over{E_0}} \times {{E_{\rm
      init}}\over{E_0}}} \geq {1 \over{2 \omega P_{\rm
        orb}}} \, .
\label{eq:threshold_nl}
\end{equation} 
There is a simple explanation for this reduction of threshold. In the
toy model, which involves the addition of a series of vectors that
have equal lengths but different orientations, if the initial vector
is placed well away from the origin, any new vector will introduce a
much larger energy shift in the mode,
\begin{eqnarray}
E_{i} - E_{i-1} & \propto & |{\bf b}_i|^2 - |{\bf b}_{i-1}|^2 
 \nonumber \\
& \approx &  2 {\bf b}_{i-1} \cdot \Delta {\bf b}_i + |\Delta
{\bf b}_i|^2 \, ,
\label{eq:bvec}
\end{eqnarray}
than if the initial vector is near the origin ($\sim |\Delta {\bf
  b}_i|^2$). This corresponds to a bigger change in the orbital
period, and therefore a larger change in the kick phase the next time
around. As a result, the threshold depends on the geometric mean of
the initial and the one-kick energy. 

Physically, the energy exchange between the orbit and the f-mode is
enhanced when there is a pre-existing large-amplitude f-mode. This
comes about because the f-mode can now perturb the orbit more
efficiently (eq. \ref{eq:orbitevolve}), thereby affecting its own driving.
In fact, the energy exchange accelerates as the f-mode gains energy.


\section{Application to secular migration}
\label{sec:secular}

I now return to the initial motivation for this work, the migration of
hot Jupiters.  We can now see how f-mode diffusion can effectively
stall the secular migration, preventing the plants from being tidally
disrupted, as well as how the orbits of these planets are subsequently
circularized.

\subsection{Stalling the secular migration}

\begin{figure} 
\includegraphics[width=0.43\textwidth,trim=0 170 50 110,clip=]{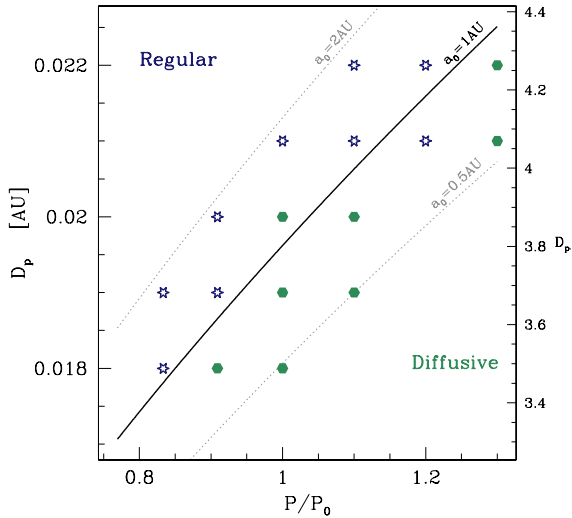} 
\caption{The critical pericentre distance below which f-mode diffusion
  will occur, plotted here as a function of f-mode period (normalized
  by $P_0 = 1.04\times 10^4\s$) and applies when $E_{\rm init}=0$. The
  solid line represents the analytical scaling (
  eq. \ref{eq:threshold}) and the dots are results of orbit
  integration (filled green circles for diffusive and open stars for
  regular). The planet's has an initial orbit of $a_0 = 1$ AU, and the
  two grey curves indicate how the critical distance is expected to
  change when it instead starts at $a_0 = 2$ and $0.5$ AU,
  respectively.  The right axis is in units of tidal radius, where $r_t =
  (M_*/M_p)^{1/3} R_p$.}
\label{fig:threshold}
\end{figure}

Consider a migrating Jovian planet with an ever decreasing pericentre
distance. Its f-mode is initially un-excited.  Combining
eqs. \refnew{eq:Klmwu}, \refnew{eq:fracE} \& \refnew{eq:kickmin}, one
finds that diffusive tidal evolution of this mode will kick in when
\begin{eqnarray}
D_p & \leq& 
 0.02 {\rm AU}
\left({{R_p}\over{1.1 R_J}}\right)^{2/21}\,
\left({a_0\over{1 {\rm AU}}}\right)^{5/42} \, 
\left({{Q_{n\ell}'}\over{0.5}}\right)^{2/21} \nonumber \\
& & \times \left({{P}\over{1.04\times 10^4\s}}\right)^{11/21}\, .
\label{eq:threshold}
\end{eqnarray}
{\wyq Notice that the $P$ here referes to the period of the prograde
  mode.}
The numerical version of this expression in plotted in
Fig. \ref{fig:threshold}, together with supporting evidences from our
numerical integrations.

This is one of our key result. The critical $D_p$ has a very weak
dependences on almost all parameters, and a weak dependence on the
mode period. 
To make explicit the dependence on planet properties, one writes
$P
\approx P_0/(1 + P_0/P_{\rm spin})$, where $P_{\rm spin}$ is the
planet spin period {\wyq (positive if spin aligns with the orbit)}, 
  $P_0$ is the $\ell=2, m=0$ f-mode period and it scales with bulk
  planet properties as $P_0 \approx 1.04\times 10^4 \s
  (M_p/M_J)^{-0.48} (R_p/1.1 R_J)^{2.1}$ \citep{Burrows}.  As such,
  eq. \refnew{eq:threshold}, measured in unit of tidal radius $r_t =
  R_p (M_*/M_p)^{1/3}$, becomes
\begin{eqnarray}
{{D_p}\over{r_t}} & \approx & {3.8} \times (1+{
        P_0/P_{\rm spin}})^{-0.52}\, 
\left({a_0\over{1 {\rm AU}}}\right)^{0.12} 
\nonumber \\
& & \times 
\left({{R_p}\over{1.1 R_J}}\right)^{0.19}\,
\left({{M_p}\over{M_J}}\right)^{0.08} \, 
\left({{Q_{n\ell}'}\over{0.5}}\right)^{0.09} \, ,
\label{eq:thresholdrt}
\end{eqnarray}
The critical $D_p$ falls within a narrow range around $4$ tidal radii
(also see Fig. \ref{fig:threshold}). If one takes the definition of the
Roche radius to be $R_{\rm Roche} = 2.44 r_t$, then the critical $D_p
\sim 1.6 R_{\rm Roche}$.

When the f-mode starts diffusing, orbital energy is quickly
transferred to the internal oscillations and the orbit decays.  This
effectively decouples the planet from secular forcing by its
companions.
The rate of decay depends on $\Delta E_{\rm kick}$.  The top example
in Fig. \ref{fig:example} shows that it takes $\sim 400$ yrs for the
orbit to decay from $1$ to $0.8$ AU. Since the strength of secular
coupling goes down with $a$ ($\propto a^{3}$ for quadrupole coupling),
and because secular forcing depends sensitively on the concordances
among different secular frequencies (which depend on $a$ nonlinearly),
such an orbital decay substantially reduces the secular forcing and
prevents the planet orbit from getting even closer to the star. 
To further strengthen this point, one notes that since the one-kick
energy scales with $D_p$ as $D_p^{-21}$, a minute drop in $D_p$ is
sufficient to overcome any strength of secular forcing, even if the
diffusion is initially too slow to stall the migration. The planet is
safely parked {\wyq around that predicted in eq. \refnew{eq:thresholdrt}}.

The f-mode also introduces an apsidal advance that can help to
decouple the planet from secular forcing. It is found numerically that
the precession rate is a few times higher than that predicted using
equilbrium tide theories \citep{Sterne1939,Smeyers91}. But it has a
weaker dependence on $D_p$ as $D_p^{-5}$, so it helps to stall
migration with weak secular forcing (e.g., the case of HD 80606), but
the orbital decay is a more fail-proof mechanism.

\begin{figure} \includegraphics[width=0.48\textwidth,trim=20 175 60
  95,clip=]{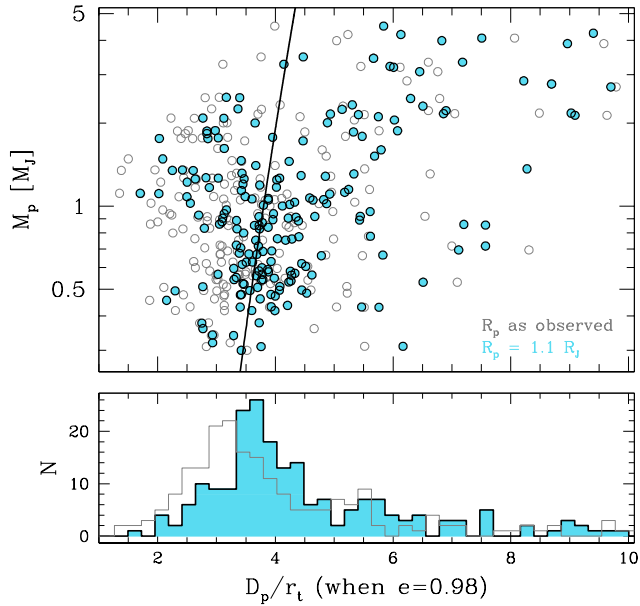}
  \caption{The stalling distances of hot Jupiters with known radii and
    masses. In the top panel, the horizontal axis is the
periastron distance {\w when $e=0.98$,}
plotted in unit of the tidal radius, while the vertical axis is planet
mass (in unit of Jupiter mass).  The grey open circles are obtained
using planets' current observed radii, while the blue solid ones are
obtained by assuming $R_p = 1.1 R_J$.  The {\w thick black line} is
the theoretical threshold for f-mode diffusion
(eq. \ref{eq:thresholdrt}), assuming $P=1.04\times 10^4\s$ and zero
spin. While low-mass hot Jupiters cluster around this {\w prediction},
high mass ones are much more spread out.  The bottom panel shows the
corresponding histograms. There is a strong {\w pile-up just below} 4
$r_t$, {\w as predicted, if planets are indeed $\sim 1.1 R_J$ in
  radius when they were migrated.}}
\label{fig:tidal-radius-m}
\end{figure}

One can now compare the critical distance against the current
positions of known hot Jupiters. Most of them have now near zero
eccentricities, so at high eccentricities, they should satisfy $D_p
(e\approx 1) = a (e \approx 0)/2$. I plot these values against planet
masses in Fig. \ref{fig:tidal-radius-m}, and find that most the
observed hot Jupiters satisfy $D_p${\wyq $(e\approx 1)$} $\sim 3 r_t$
if we adopt their current (inflated) radii, and $D_p \sim 4 r_t$ if we
assume instead that during migration, their radii $R_p = 1.1 R_J$.

{\wyq This is expected. The} progenitors of hot Jupiter likely possess
the same orbital distribution as the cold Jupiters found today by
radial velocity surveys, namely, a precipitous rise just outside
$a=1$AU and a gradual fall further out. So {\wyq
  eq. \refnew{eq:thresholdrt} predicts $D_p (e\approx 1) \sim 3.8 r_t$
  for non- or slowly-spinning planets, and slightly smaller values for
  rapidly spinning planets.}  Moreover, there should be a sharp pile-up
  around this value due to the weak dependence of critical $D_p$ on
  all relevant parameters. As Fig. \ref{fig:tidal-radius-m} shows, the
  observed spread around $4 r_t$ is indeed narrow for planets less
  massive than Jupiter, but appears to be much broader for higher mass
  planets. {\wyq The model here could not account for this latter
    behaviour but I note that for these higher mass planets, tidal
    excitation in stars may become more relevant \citep[see,
    e.g.][]{IvanovPapaloizou,BarkerOgilvie}.}

\subsection{Towards Orbital Circularization}
\label{subsec:circorbit}

Now consider the Jupiter after its secular migration has been
stalled. It now {\wyq resides on} a high eccentricity orbit with $D_p$
near the original threshold, {\wyq largely independent of the secular
  forcings. } As the f-mode gains energy, the nonlinear criterion
(eq. \ref{eq:threshold_nl}) takes hold. This {\wyq now facilitates} the
eventual circularization -- as the orbit decays and $D_p$ gradually
rises, the one-kick energy drops precipitously. The nonlinear
threshold reduces the one-kick energy required for diffusion and
allows the planet to continue on its way to circularization.

We define another energy scale for the f-mode, $E_1$. Let $\Delta r$
be its surface radial displacement at the equator, we define 
\be E_1 =
E (\Delta r = R_p) \sim {0.1 {{GM_p^2}\over{R_p}}} \sim 3 \times
10^{42} \erg\, .
\label{eq:E1}
\ee
For comparison, the orbital energy at $a=1$ AU is $9\times 10^{42}
\erg$.  Here, I restrict $E \leq E_1$ and discuss the nonlinear
evolution of the f-mode in a later section. 

\begin{figure} 
\includegraphics[width=0.36\textwidth,trim=0 150 50 70,clip=]{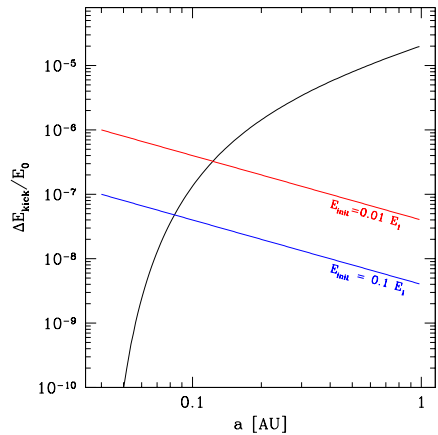} 
\caption{The reach of diffusive tidal evolution. Here, starting from
  $D_p = 0.02$ AU and $a=1$ AU, the planet moves inward on a
  trajectory of constant angular momentum (horizontal axis is the
  semi-major axis, in AU). The vertical axis is the one-kick energy it
  is expected to receive along this trajectory and it drops
  precipitously during the evolution (black curve), as a result of a
  slight increase in $D_p$. This energy is normalized by the local
  orbital energy ($E_0= G M_* M_p/2a$), and $P = P_0$.  The
  two coloured curves indicate the minimum one-kick energy required to
  cause tidal diffusion, when the f-mode energy takes on two initial
  values ($0.01$ or $0.1 E_1$, where $E_1$ corresponds to unity radial
  displacement at the surface). In both cases, the diffusive evolution
  can be sustained inward till $a \sim 0.1$ AU. Here, the orbit
  integral $K_{\ell m}$ is calculated as in eq. \refnew{eq:dong}. }
\label{fig:threshold_nl}
\end{figure}

Let us assume that the earlier evolution has endowed the f-mode with a
non-zero initial energy that is a fraction of $E_1$. As is shown in
Fig. \ref{fig:threshold_nl}, the minimum distance the planet can reach
via diffusive evolution depends on this fraction.  Starting from a
high-eccentricity orbit with $D_p = 0.02$ AU, if the energy
fraction is $10^{-2}$, the planet can continue to experience diffusive
tidal evolution (satisfying eq. \ref{eq:threshold_nl}) until its orbit
has shrunk to $a=0.13$ AU ($e=0.83$); and if the fraction is raised to
$10^{-1}$, the evolution can proceed further till $a=0.09$AU
($e=0.73$).


The change in orbital energy between the above final ($a\sim 0.1$ AU)
and initial orbits ($a=1$ AU) exceeds the binding energy of the planet
by a factor of {\wyq a} few. Can a single f-mode carry the planet inward for
such a large distance?

\begin{figure}
\includegraphics[width=0.45\textwidth,trim=0 220 10 70,clip=]{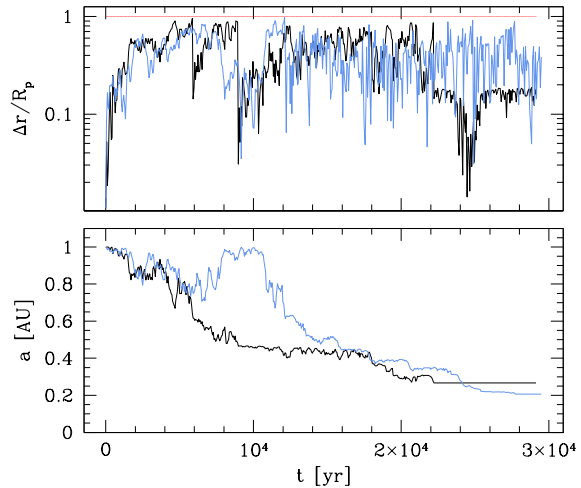}
\caption{Diffusive tidal evolution involving only one f-mode. A Jovian
  planet (with $P = P_0$) is initially placed at $a=1$AU and $D_p =
  0.02$ AU ($e=0.98$). The top panel shows the fractional radial
  displacement, measured at the surface of the planet, and the bottom
  panel the semi-major axis of the orbit.  Whenever the diffusion has
  grown the f-mode to unity amplitude (red line), I prescribe the
  code to remove $99\%$ (the solid black curve, or $90\%$ for the
  dashed blue curve) of the mode energy instantaneously, down to $1\%
  E_1$ (or $10\% E_1$). After $\sim 10^4$ yrs, the planet orbit has
  tidally circularized to $a\sim 0.2$. After this point, diffusion is
  inhibited (Fig. \ref{fig:threshold_nl}) and the tidal evolution
  stalls. }
\label{fig:mock_finiteamp}
\end{figure}

In Fig. \ref{fig:mock_finiteamp}, I present a scenario to illustrate
how I believe this is accomplished. The planet is initially placed at
an orbit with $a = 1$AU, $D_p=0.02$AU ($e=0.98$). Its f-mode starts
random-walking and this continues until its amplitude has reached
unity. At this point, nonlinearity is important (\S
\ref{subsec:nonlinear}). Here, I simply specify that the mode energy
be instantaneously removed, leaving a small residual energy, and that
there be no changes in the planet's properties (radius, mass, spin
rate). Thanks to the residual energy, the f-mode remains diffusive,
and is soon undergoing another nonlinear damping.  Within an
astronomically short time (a few $10^4$ yrs), the orbit has decayed to
$a \sim 0.2$ AU, or an eccentricity of $e \sim 0.9$. Evolution
practically stalls after reaching this point.

These stalling distances lie twice above our analytical predictions
($a=0.1$ AU, Fig. \ref{fig:threshold_nl}). The reason may be observed
in the right-hand panel of Fig. \ref{fig:Klmfits-new} -- by the time
the planet has migrated to these distances, the parabolic
approximation for the orbit integral is no longer valid, it is instead
dominated by a series of resonances. Our simple treatment should fail
in this regime.


If our scenario is correct, there are two immediate implications,
one relates to the effective tidal $Q$ number, the other relates to
the observed absence of very high eccentricity planets.

The amount of energy transferred to the f-mode per orbit is $\sim
\sqrt{E \Delta E_{\rm kick}}$, where $E$ is the mode energy
(eq. \ref{eq:bvec}).
To re-cast our results using the so-called tidal-quality factor ($Q$),
we invoke the following expression for tidal orbital decay, valid for
high eccentricity orbits \citep{MacDonald,GoldreichMurray},
\begin{equation}
{1\over a}{{da}\over{dt}} = - {{21}\over{64}} {n\over\mu} {{a
    R_p^5}\over{D_p^6}} {{k_2}\over{Q}}\, ,
\label{eq:defineQ}
\end{equation}
where $n$ is the orbital frequency, $\mu = M_p/M_*$, and $k_2$ is the
tidal Love number which I take to be $0.3$. Combined with
eq. \refnew{eq:Ekick}, eq. \refnew{eq:Klmwu} and eq. \refnew{eq:E1},
this yields
\begin{eqnarray}
Q & = & {{42\pi}\over{64}} {1\over\mu} {{a
    R_p^5}\over{D_p^6}} {{k_2 E_0}\over{\sqrt{E
      \Delta E_{\rm kick}}}}\nonumber \\
& \approx & 73 \sqrt{{E_1}\over{E}}
\left( {{R_p^3}\over{M_p}} \times {{M_*}\over{D_p^3}}\right)^{3/2}
\left({{K_{\ell m}}\over{0.006 T_{\rm peri}}}\right)^{-1}
\nonumber \\
& \approx & 0.5 \times \left({{E\over{10\%
      E_1}}}\right)^{-1/2} \left({{D_p}\over{0.02 {\rm
        AU}}}\right)^{4.5} \left({{P}\over{1.04\times 10^4\s}}\right)^{-6} \, ,
\label{eq:Qeq}
\end{eqnarray}
where I have scaled the f-mode energy $E$ by typical values observed
in Fig. \ref{fig:mock_finiteamp}. 

Such a small $Q$ factor differs from the common conception that $Q
\sim 10^5$ for Jovian planets, a conception that comes from
constraints on the Gallilean satellites which move on nearly circular
orbits \citep{GoldreichSoter}. It makes sense, however, on hind sight:
the f-mode is the equilibrium tide, or most of it; and the f-mode
diffusion moves energy in such a way that of order the equilibrium
tide energy is effectively absorbed by the planet after every passage,
{\wyq though} the true dissipation (nonlinear breaking) {\wyq really}
only sets in {\wyq when the mode is} at unity amplitude.

Fig. \ref{fig:mock_finiteamp} shows that, starting from an eccentricity
of $e =0.98$, a Jovian planet can circularize its orbit to $e < 0.9$
within a few $10^4$ years. Let this timescale be $10^5 \yrs$ to be on
the conservative side. One can estimate the number of super-eccentric
planets ($e > 0.9$) one expects in the {\it Kepler} sample of
$200,000$ stars. First,  assume that about $1\%$ of the stars could
eventually own a hot Jupiter, based on the observed frequency of hot
Jupiters. If the event that makes them occur relatively uniformly
during the stars' lifetimes ($\sim 5$ Gyrs), the number of
super-eccentric Jupiters should be \be N_{\rm e > 0.9} \approx 200,000
\times 1\% \times {{10^5 \yrs}\over{5 \times 10^9 \yrs}} \sim 0.04\, .
\label{eq:supere}
\ee 
This value is further reduced when one considers the geometric
probability of a planetary transit. 

This explains the observed deficit of such planets \citep{Dawson},
despite the arguments presented in \citet{Socrates}. Tidal dissipation
at very high eccentricity proceeds efficiently, likely far more
efficient than when the planet is less eccentric. About the latter we
still have no good first-principle theory.


\section{Miscellaneous}

Here, I justify a number of assumptions in our model, discuss the
nonlinear behaviour of the f-mode, the impacts of nonlinear damping on
the planet, and compare our results with previous studies.

\subsection{Assumptions}

I only consider one $\ell=2, m=-2$ f-mode. The other modes contribute
at the percent level and can be safely ignored \citep[also
see][]{Mardling}. This is due to a number of factors: the tidal
potential drops off for higher multiples ($\ell$); the tidal integral,
$Q'_{n\ell}$, drops off with the mode's radial order; the orbit
integral, $K_{\ell m}$, drops off steeply with decreasing mode
periods. Relatedly, when the f-mode is strongly excited, it could help
make the other modes to go stochastic, However, since the one-kick
energy is the largest for the f-mode, it still diffuses the
fastest. As a result, it dominates the orbital evolution.


I ignore linear dissipation on the f-mode. The dominant viscosity in
a fully convecting planet is turbulent viscosity. But since the
convection over-turn time ($\sim$ yr) is some $4000$ times longer than
the f-mode period, the effective damping time is $\sim 4000^2 \sim 10^7$
yrs \citep{GoldreichNicholson}. This is far longer than the longest
timescale of interest here, $\sim 10^4$ yrs.

In our model, I assume the planet bulk properties (radius, mass, spin
rate) remain constant throughout the evolution, despite the repeated
nonlinear breaking of the f-mode. I give arguments to support this in
\S \ref{subsec:nonlinear}.

I have ignored tidal response in the star. To estimate the relative
importance of the stellar f-mode, let us swap the subscripts for the
star and the planet in eq. \refnew{eq:Ekick}, to find that, for the
same body density and a similar mode period, the f-mode energy gain in
the star is roughly $R_p/R_* \sim 1/10 $ times of that in the planet,
{\wyq for a $1-M_J$ planet. This ratio is less extreme for more
  massive planets.} Solar gravity-modes, on the other hand, may in
fact be more important than the f-mode -- they are coupled to the
tidal potential less strongly (smaller $Q'_{n\ell}$), but their lower
frequencies may enhance $K_{\ell m}$ {\wyq
  \citep{IvanovPapaloizou}}. This study falls short of investigating
this {\wyq and this may explain our failure to reproduce the orbits
  for the high mass planets in Fig. \ref{fig:tidal-radius-m}.}


\subsection{Nonlinear Evolution}
\label{subsec:nonlinear}

The gravitational binding energy of a Jovian planet,
\be
{{G M_J^2}\over{R_J}} \sim 4\times 10^{43} \erg\, ,
\label{eq:Egrav}
\ee is comparable to the magnitude of its orbital energy at a few AU,
\be E_0 = |E_{\rm orb}| = \left| -{{GM_\odot M_J}\over{2 a}}\right| \sim
10^{43} \erg\, \times \left({a\over{1 {\rm AU}}}\right)^{-1}\, .
\label{eq:Eorb1}
\ee This suggests the importance of internal oscillations in modifying
the orbit. It also suggests that the internal oscillations can acquire
enough energy to go nonlinear. The evolution
subsequent to this is uncertain. I give my educated guess below. 

Can f-mode be saturated at a very low amplitude, {\wyq much below
  unity}?  Let us consider mode coupling to transport energy out of
the f-mode.  In a fully convective planet like Jupiter, gravity-modes
do not exist, and inertial-modes lie too low in frequency (unless the
planet is near critical spin). As a result, the only 3-mode coupling
for the f-mode involves it (twice) and another f- or p-modes at twice
its frequency. However, the sparse spectrum of these latter modes
implies that this coupling is typically far from resonance, and the
energy transfer rate is limited unless the f-mode has reached order
unity amplitude ($\Delta r \sim R_p$), by which time 4-mode coupling
may be just as important as 3-mode coupling.  This is analogous to the
situation in Cepheids and RR Lyrae pulsation where the over-stable
{\wyq f-mode} pulsation grows to unity amplitudes.

What happens when the f-mode reaches order unity amplitude? Both mode
coupling and wave breaking are possibilites to convert its energy into
heat. \citet{KumarGoodman} studied the 3-mode coupling by solving the
equation of fluid motion, with the f-mode acting as an inhomogeneous
forcing term at twice its own frequency. They found that energy is
taken out of the f-mode at a rate comparable to its own frequency when
$E \sim E_1$. Moreover, they found that the forced response peaks near
the surface ($< 100$ bar), and may itself be prone to further
nonlinear damping.  Another study of note is that by \citet{Kastaun}.
Using a general relativistic hydro-code to study a star undergoing a
large f-mode oscillation, in the context of rapidly spinning neutron
stars, they reported that when the f-mode has unity amplitude, the
stellar surface is gradually distorted away from a sinusoidal
waveform, leading to steepening and wave breaking, analogous to the
breaking of ocean waves.
In either of these calculations, the f-mode energy excites near
surface phenomenon, and energy dissipation occurs near
the surface, possibly within a few scale heights of the photosphere.

When the f-mode is (nonlinearly) dissipated, what happens to the
planet's structure and spin?  The rapid dissipation of the f-mode in a
few oscillation timescale gives rise to a luminosity of $L \sim
10^{37} \erg/\s$, outshining even the host star.
This is much beyond the planet's Eddington luminosity ($\sim 10^{35}
\erg/\s$), and is much more than can be carried out by convection or
radiative diffusion.
As a result, it must lead to envelope expansion
and mass loss.
But since all the nonlinear dissipation occurs at the low density,
superficial region of the planet, 
the mass loss rate is not significant, and the heated layer can cool
within a short amount of time. More importantly, it is hard to
transport entropy up the temperature gradient toward the planet
interior, so one expects little impact on the internal entropy.  The
radius of the planet at constant entropy, on the other hand, goes as
$R \propto M^0$, as the polytropic index $n \sim 1$ under the combined
effect of electron degeneracy and Coulomb force. So, to first order,
one can assume that the planet's radius hardly changes. The spin
evolution {\w should also be impacted by the mode of energy
  dissipation. In the case where f-mode dissipation only occurs at the
  surface, the angular momentum it carries will also likely be lost to
  the expanding envelope and is quickly removed. In this case, there
  is little spin evolution in the planet interior. Both these
  considerations justify, to some degree, my simplification of keeping
  a constant f-mode period.}

After nonlinear damping is finished, it is likely that the f-mode
still retains a fraction of its former energy. In particular,
oscillations in the central region is still very linear ($\xi_r \ll
R_p$) and may not be completely removed. This motivates us, 
in Fig. \ref{fig:mock_finiteamp}, to assume that a small fraction
remains to seed the subsequent random-walk. 


Would the now severely distended planet undergo tidal disruption?
\citet{SridharTremaine} obtained that the threshold for tidal
shredding of an incompressible, homogeneous sphere on parabolic orbit
lies at $1.69 r_t$.
At a distance of $4 r_t$, the planet is safe from tidal disruption
even if one assumes its radius is diluted by a factor of 2 by
pulsation. 

In summary, as the planet's orbit decays, the f-mode repeatedly breaks
near the planetary surface, depositing energy and angular momentum in
the top layers, possibly driving a wind. However, the bulk of the
planet may feel little impact. One does not expect the planet to be
thermally ablated. But detailed investigation is required to assess
the damage.

{\w An additional concern arises when the mode amplitude becomes very
  large. Mode period may be shifted nonlinearly and the pulse shape
  may become anharmonic. We have not included these into our
  consideration and they may impact the long-term evolution.}





\subsection{Comparing with previous studies}

{\w \citet{Mardling} studied the chaotic diffusion under extreme tidal
  forcing. She used numerical simulations to delineate the boundary
  between chaos and regular behaviour, for a system of two equal mass,
  $n=1.5$ polytropes. In particular, she showed that a highly
  eccentric orbit at $e=0.98$ (the fiducial case considered here) can
  kick start tidal diffusion when $D_p \leq 4.2 r_t$, slightly larger
  than our prediction of $3.8 r_t$. The difference may be due to our
  different assumptions on the model structure.
  Furthermore, she demonstrated numerically that a non-zero initial
  energy can boost diffusion (her Fig. 15). We provide an explanation 
  for this effect, as well as present a quantitative criterion.}

\citet{IvanovPapaloizou} {\w was the first to explain the physics
  behind the f-mode random-walk, using analytical arguments and a
  toy-model. Their study is the closest to ours as they also focussed
  on a Jovian planet around a solar-type star.}  Their central result
is their eq. (107) where they showed that, for a certain value of
$D_p$, there is a minimum $a_{\rm st}$ above which the tidal dynamics
is stochastic. Substituting our expression for $K_{\ell m}$ into their
notation, one obtains that $a_{\rm st} \propto D_p^{-42/5} P^{-22/5}$,
reproducing the scalings in our eq. \refnew{eq:threshold}, and with a
similar normalization.

 A study recently appeared while I was preparing the
 manuscript. \citet{Dong} adopted the 2-D mapping approach to
 investigate the same dynamics {\w as we study here}.  Their Fig. 1 is
 similar to our Fig. \ref{fig:randomkick-color}. They have also
 generalized the model to include effects of mode dissipation,
 resonances and gravity-modes. These are important for stars (the case
 they consider) but not for Jovian planets.




 {\w Lastly, \citet{PI05,IP07} considered the stochastic excitation of
   inertial modes, in lieu of f-modes studied here. They calculated
   that a couple low-order inertial modes can couple to the tidal
   potential sufficiently strongly \citep[also see Fig. 2
   in][]{Wu05b}, that they can potentially supplant the f-modes, for
   cases where the peri-centre distance is larger than the values
   considered here. This occurs because the longer periastron passage
   time weakens the orbit integral for f-modes, while inertial modes
   may not suffer as much, if one assumes that the rotation frequency
   remains comparable to $1/T_{\rm peri}$. This may be another venue
   for tidal circularization.}

\section{Conclusion}

In this work, I use both direct numerical integration and a toy-model
to investigate the tidal evolution of a Jovian planet on a highly
eccentric orbit around its host star. The findings here allow us to
overcome three of the theoretical weaknesses in dynamical migration
for hot Jupiters, and boost the overall prospects for dynamical migration.

I show that, when the planet's pericentre dips below $4$ tidal radii,
one of its f-modes starts gaining energy stochastically. Because there
is more phase space for the f-mode at high energy, and because the
orbit and the f-mode try to reach energy equi-partition, the f-mode
diffusively grows towards unity amplitude. The growth of the f-mode is
accompanied by the decay of the orbit. So a Jupiter that is secularly
perturbed to high eccentricity will be stalled and dynamically
decoupled from its perturbers when its pericentre distance reaches
$\sim 4$ tidal radii. They are safely parked where they are observed
today, without suffering the fate of tidal disruption.

One of the 'accepted' examples for secular migration is HD 80606
\citep{WuMurray}. At its current orbit of $D_p = 0.03$ AU and
$e=0.983$, it should never have crossed inward of $D_p = 0.029$ AU, or
$11 r_t$ ($M_p = 4 M_J, R_p = 0.9 R_J$). This is too far for f-mode
diffusion.
However, if this planet is indeed migrated inward by its remote
stellar companion ($\sim$ a thousand AU) as suggested, {\wyq the weak
  secular perturbation from the companion} can be easily stalled at
the observed distance by tidal and secular precessions
\citep{WuMurray}, without the need to invoke diffusive tidal
evolution.

One of the new insights in this work is that mode diffusion can occur
at a lower threshold when the f-mode has some non-zero energy to start
with {\wyq \citep[also see][]{Mardling}}. This insight is important
for transporting the planet all the way from a few AU to a small
fraction of an AU. In our simulations, {\wyq we model this by}
assuming that whenever the f-mode nonlinearly damps, a fraction of the
initial energy is retained to seed the next episode of random-walk.
These simulations show that, within a few $10^4$ years, the planet
drops its eccentricity from near unity to $0.9$. We therefore do not
expect to see any super-eccentric Jupiters ($e > 0.9$) among the
$200,000$ stars observed by the {\it Kepler} mission. More strikingly,
the process discussed here achieves an effective tidal $Q \sim 1$, in
an otherwise invisid planet. This helps explain how the planet
dissipates the tide, without invoking any ad hoc weak friction.

To emphasize the last point, I note that although the planet's orbit
decays and circularizes due to f-mode diffusion, these changes are
temporary and occur without any explicit dissipation in the system. It
is only when the f-mode is nonlinearly dissipated, either through
mode-coupling or wave breaking, these orbital changes are perpetuated
and tidal circularization becomes time irreversible.


Our story fails, however, after $a\sim 0.2$ AU. What mechanism is
capable of further circularizing the planet to the nearly zero
eccentricity we see today?  Can the residual energy in the f-mode be
again useful? If the tidal process proceeds much more slowly in the
later stage, are there any observational consequences (e.g., warm
jupiters from stalled circularization)?

Lastly, the fate of a tidal planet under the massive amount of energy
deposition need not be dire. I argue that f-mode dissipation occurs
exclusively near the surface. As the f-mode energy is converted into
heat, this should lead to envelope expulsion but should keep the
planet interior largely intact.

\acknowledgments {This research was started while I was visiting IAS
  and I thank Scott Tremaine and the gang (Tejaswi Nerella, Liang Dai,
  Morgan MacLeod, Adrian Hamers) for a warm hospitality and a
  stimulating environment. I also thank Yoram Lithwick, Cristobal
  Petrovich and Norm Murray for discussions. {\w I am grateful to
    Pavel Ivanov for a critical review that corrected a mistake in my
    earlier draft}.  Lastly, I am grateful to Peter Goldreich for
  my perennial interest in tidal dissipation. }

\bibliographystyle{apj}
\bibliography{ref}

\end{document}